\documentclass[aps,pra,twocolumn,showpacs,amssymb,amsmath,amsfonts,superscriptaddress,floatfix,nofootinbib,groupedaddress]{revtex4-2}
\usepackage{graphicx}
\usepackage{mathtools}
\usepackage{perpage} 
\usepackage{amsmath,amssymb}
\MakePerPage{footnote}
\usepackage{color}
\usepackage{braket}
\usepackage{subfigure}

\begin{document}
	
	\title {Quantum Multiplication Algorithm Based on the Convolution Theorem}
	\author{Mehdi Ramezani, Morteza Nikaeen, Farnaz Farman, Seyed Mahmoud Ashrafi and Alireza Bahrampour}
	\affiliation{Department of Physics, Sharif University of Technology, Tehran 14588, Iran}
	\affiliation{Centre for Quantum Engineering and Photonics Technology, Sharif University of Technology, Tehran 14588, Iran}
	
	\date{\today}
	
	\begin{abstract}
		The problem of efficient multiplication of large numbers has been a long-standing challenge in classical computation and has been extensively studied for centuries. It appears that the existing classical algorithms are close to their theoretical limit and offer little room for further enhancement. However, with the advent of quantum computers and the need for quantum algorithms that can perform multiplication on quantum hardware, a new paradigm emerges. In this paper, inspired by convolution theorem and quantum amplitude amplification paradigm we propose a quantum algorithms for integer multiplication with time complexity $O(\sqrt{n}\log^2 n)$ which outperforms the best-known classical algorithm, the Harvey algorithm with time complexity of $O(n \log n)$. Unlike the Harvey algorithm, our algorithm does not have the restriction of being applicable solely to extremely large numbers, making it a versatile choice for a wide range of integer multiplication tasks. The paper also reviews the history and development of classical multiplication algorithms and motivates us to explore how quantum resources can provide new perspectives and possibilities for this fundamental problem.
	\end{abstract}
	
	
	\maketitle
	
	
	\section{Introduction}
	Efficient computation of integer multiplication, as one of the most elementary mathematical operations, is crucial in many fields including computer science and engineering. The development of multiplication algorithms has been a fascinating journey that has spanned centuries and continents. From the ancient Egyptians to modern-day mathematicians, people have been searching for more efficient ways to perform multiplication. Today, we have a range of algorithms that allow us to perform complex calculations quickly and accurately.
	
	Grade-school algorithm, also known as the standard algorithm, is the most basic and widely taught multiplication algorithm. However, its time complexity $O(n^2)$, i.e., the number of single-bit arithmetic operations necessary to multiply two $n$-bit integers, limits its practicality for larger numbers. Although many methods were invented such as Egyptian multiplication and lattice multiplication, their time complexity was also limited to $O(n^2)$.
	
	For years, it was believed that the time complexity of multiplication algorithms could not be improved beyond $O(n^2)$. This changed in the 1960s with the introduction of the breakthrough Karatsuba algorithm \cite{karatsuba1962multiplication}, which revolutionized multiplication algorithms by reducing the time complexity to $O({n^{{{\log }_2}3}})$. This is a divide-and-conquer algorithm that splits the numbers into two smaller parts and recursively multiplies them until the smallest parts can be multiplied directly. Because of the overhead of recursion, Karatsuba's multiplication is slower than standard multiplication for small values of $n$. However, it is asymptotically faster than standard multiplication becoming a cornerstone of modern multiplication algorithms and has paved the way for even faster algorithms.
	
	It didn't take long for Karatsuba's idea to lead to a faster algorithm known as the Toom-Cook algorithm \cite{toom1963complexity}. It is an extension of the Karatsuba algorithm that splits the numbers into more than two parts. Due to the growth of the overhead from additions and digit management, this algorithm has a higher computational complexity than the Karatsuba algorithm but can be faster for very large numbers, typically used for intermediate-size multiplications. For example, the time complexity of Toom-3, which splits the numbers into three parts, is $O({n^{{{\log }_3}5}})$.
	
	Karatsuba algorithm is by no means the end of the line for multiplication algorithms but emerged as the beginning of modern multiplication algorithms.
	A further step in modern multiplication algorithms was taken by Arnold Schönhage and Volker Strassen in 1971 \cite{schonhage1971schnelle},\cite{schonhage1977schnelle}. Their work involves representing each number as a polynomial, where the coefficients of the polynomial correspond to the digits of the number. Convolving the vectors of polynomial coefficients is equivalent to multiplying the two polynomials. The algorithm then uses the fast Fourier transform (FFT) and convolution theorem to compute the product of these polynomials, which gives the product of the original numbers. This approach exploits the efficiency of the FFT algorithm to reduce the time complexity of multiplication. The FFT and convolution theorem provide a way to compute the product of two extremely large numbers which asymptotically has much lower computational complexity than the Toom-Cook algorithm. The original algorithm performs the discrete Fourier transform over complex fields leading to a time complexity of $O(n{\log _2}n)$. However it requires the use of complex numbers and floating-point arithmetic, and as such, it requires a significant amount of memory to store intermediate results due to the numerous calculations involved. Therefore, implementing the algorithm to get error-free outcomes is not practical even in modern classical computer architectures. To overcome this difficulty, they modified the algorithm by changing the field over which the FFT is performed from the complex field to the finite field, i.e., the Galois field. This resulted in error-free outcomes with a trade-off time complexity of $O(n{\log _2}n{\log _2}{\log _2}n)$.
	
	$36$ years later, Martin Fürer improves the asymptotic complexity of the multiplication to $n{\log _2}{2^{O(\log *n)}}$ using Fourier transforms over complex numbers but with a different divide-and-conquer pattern than the one of Strassen Algorithm, where $log*$ denotes the iterated logarithm \cite{furer2007faster}.
	
	Efforts continued to provide new algorithms to reduce the time complexity of multiplication problems. Strassen's conjecture that the final time complexity of the multiplication algorithm is $O(n\log_2 n)$ has been a guiding principle for many researchers. In 2019, David Harvey achieved this feat by discovering of an $O(n \log_2 n)$ multiplication algorithm, which is believed to be close to the optimal solution for this problem \cite{harvey2021integer}. But the key aspect to note about this algorithm is that it can only be used for extremely large numbers with a minimum of $2^{1729^{12}}$ bits!
	
	Overall, each multiplication algorithm has its own advantages and disadvantages, and the choice of algorithm depends on the size and type of the input numbers, the available resources, and the desired level of accuracy and hardware limitations. Each of these algorithms may be more or less appropriate depending on the specific use case. 
	
	As one of the most basic mathematical operations, multiplication naturally is employed in various quantum algorithms. Therefore, devising and implementing efficient quantum multiplication algorithms that can be run in quantum computers is a crucial problem. A natural question then arises as to whether the quantum paradigm can give us quantum algorithms with some advantages over classical algorithms for multiplication problems. The majority of quantum algorithms introduced for arithmetic operations, like multiplication, often amount to quantum implementations of classical algorithms, lacking a substantial computational advantage \cite{parent2017improved,gidney2019asymptotically,ruiz2017quantum,nie2023quantum}. As an illustration, in reference \cite{ruiz2017quantum}, a technique for integer multiplication utilizing the quantum Fourier transform is put forth, albeit with a gate count of $O(n^{3})$. Furthermore, in reference \cite{nie2023quantum}, a quantum adaptation of Strassen's algorithm is introduced, offering comparable time complexity to its classical counterpart.
	
	Considering the quantum Fourier transform's (QFT) superiority compared to its classical counterpart and recognizing that qubits, unlike classical bits, can store complex numbers without compromising precision, we derived inspiration from the convolution theorem to introduce a quantum multiplication algorithm offering computational advantages. In the beginning, it is crucial to reconstruct the quantum version of the convolution theorem, appropriate for quantum resources. In doing so, in the first step, we need to encode the binary vectors of polynomial coefficients into qubits. Surprisingly it turns out that for encoding vector corresponding to an $n$-bit number, we only need $\log_2 n$ qubits. This is the reduction of space complexity of the algorithm, the first advantage we exploit from quantum resources. The intermediate step is to implement the QFT circuit of these vectors, leading to a reduction of the time complexity of the algorithm, which is the second advantage exploited by quantum resources. The final step involves building a quantum circuit for the element-wise product of two quantum vector states corresponding to two involved numbers. Unfortunately, there is no such deterministic quantum circuit \cite{lomont2003quantum}. So, we are forced to be satisfied with the implementation of the probabilistic version of this circuit that in turn increases the time (or equivalently space) complexity of the algorithm. The probability of success of the element-wise product circuit is $O(1/n)$ and for every successful run of the circuit, the time complexity of the remaining parts of the algorithm is found to be $O(\log_{2}^{2} n)$. Therefore, the overall time complexity of the algorithm will be $O(n\log_{2}^{2} n)$. However, it is possible to further enhance the algorithm's time complexity to $O(\sqrt{n} \log_{2}^{2} n)$ by leveraging the quantum amplitude amplification method. This technique effectively transforms the probabilistic aspects of the algorithm into nearly deterministic outcomes.
	
	The organization of the paper is as follows. In Section \ref{sec2}, we provide a detailed quantum circuit implementation of grade-school and the Karatsuba algorithms and analyze their resource requirements in terms of qubits, ancillas, gates, and circuit depth. This analysis provides a baseline for comparison with our proposed algorithm, which is presented in Section \ref{sec3}. In Section \ref{sec3}, we introduce the classical convolution theorem and demonstrate how it can be used to efficiently multiply integers. We then extend this theorem to the quantum domain, which involves three main steps: encoding integer vectors in qubits (presented in Subsection \ref{sub3c}), applying QFT, and devising a circuit for the element-wise product of Fourier transformed vectors (presented in Subsection \ref{sub3d}). We highlight the practical advantages of our proposed algorithm and present the results of our implementation in Qiskit in Subsection \ref{sub3e}. Finally, we conclude with a discussion that includes a comparison of our algorithm with the grade-school and the Karatsuba algorithms, as well as a summary of its advantages over the modern classical multiplication algorithms.
	
	
	\section{Quantum Circuits for Grade-School and Karatsuba Algorithms}\label{sec2}
	
	With the development of quantum computers and the demand for quantum algorithms that can perform multiplication on quantum hardware, it is crucial to design efficient quantum multiplication algorithms that are suitable for quantum hardware implementation. One of the objectives of various algorithms and techniques for implementing quantum multipliers is to optimize the number of quantum gates, time complexity, hardware complexity, garbage outputs, and constant inputs (ancillas). In this section, we examine the quantum circuits of two important multiplication algorithms: grade-school and Karatsuba. We characterize the various resources of these circuits, such as Depth (the number of time steps needed to execute the circuit), Cost (the total number of gates applied in the circuit), and Ancillas (the number of auxiliary qubits used in the circuit), to enable us to compare our proposed algorithm with the quantum implementations of these two important algorithms. We first explain the quantum circuit of the grade-school method, which uses quantum ANDing (QAC) and quantum Full adder (QFA) circuits as basic components. Then, we show how to use these components to implement the quantum circuit of the Karatsuba algorithm, which performs better than the grade-school method for larger numbers of qubits.
	
	\subsection{quantum circuit of grade-school multiplication algorithm}
\begin{figure}[htbp]
		\centering
		\includegraphics[width=\linewidth]{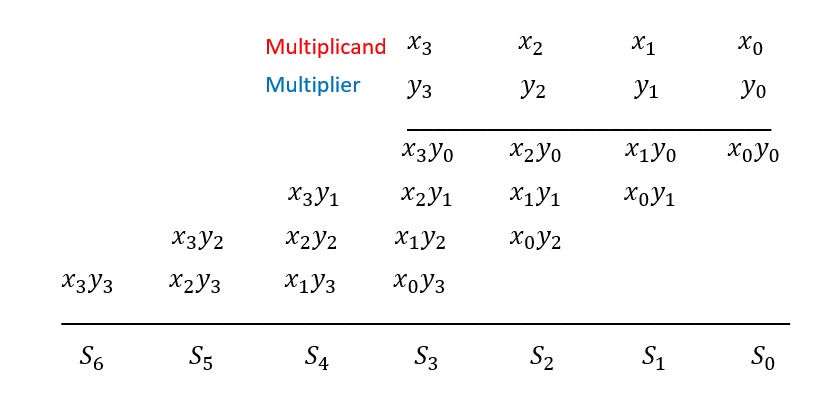}
		\caption{Grade-school multiplication of two $4$-bits numbers}
		\label{Multiply}
	\end{figure}

	To obtain a quantum circuit for the grade-school multiplication algorithm, let's consider the following example, where we multiply two 4-bit numbers, $x=x_{3}x_{2}x_{1}x_{0}$ and $y=y_{3}y_{2}y_{1}y_{0}$. In the grade-school multiplication algorithm, we multiply these numbers together in the form shown in Fig.\eqref{Multiply}. In the quantum implementation of this algorithm, we encode each bit of $x_i$ and $y_j$ in qubit $\ket{x_i}$ and $\ket{y_j}$ respectively. This encoding method is called \textit{basis encoding}.
	
	As shown in Fig.\eqref{Multiply} all multiplicand and multiplier qubits must be multiplied together and added purposefully to get the final result. So we can say all quantum multiplier circuits are composed of two sub-circuits: the quantum partial product generation circuit (PPG) and the quantum partial product addition circuit (PPA). 
	
	\subsubsection{Partial Product Generation circuit (PPG)}
	
	\begin{figure}[htbp]
		\centering
		\includegraphics[width=\linewidth]{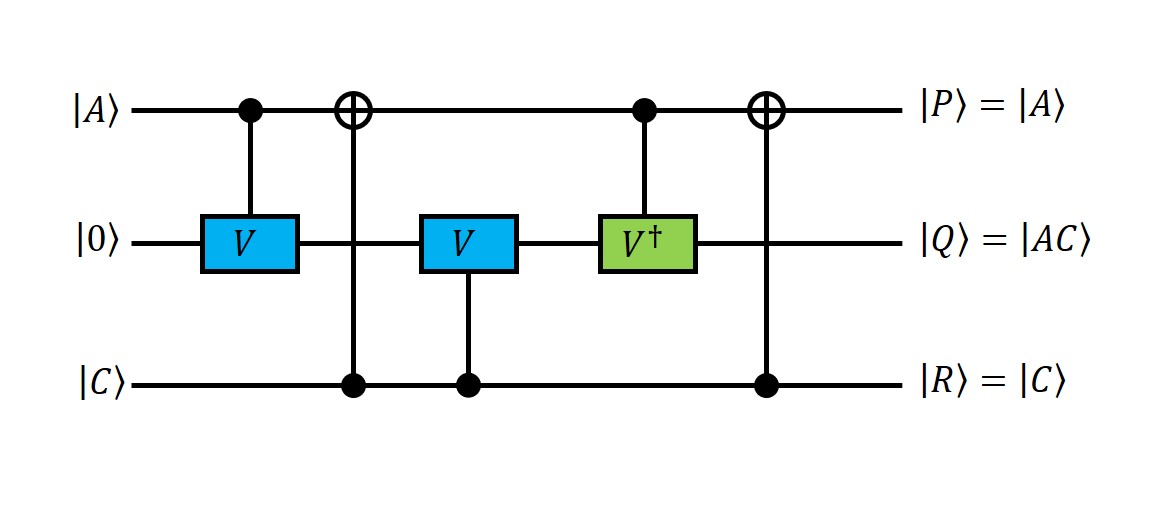}
		\caption{Schematic representation of quantum Anding circuit.}
		\label{Anding}
	\end{figure}
	
	The inputs of this circuit are the array of multiplicand and multiplier qubits. The outputs of the circuit are obtained by multiplying each of the multiplier and multiplicand qubits together. A quantum ANDing circuit (QAC) is used to multiply two qubits. The diagram of QAC is shown in Fig.\eqref{Anding}, which multiplies the input qubits  A and C and shows the result in output qubit Q. In this figure $V:=\sqrt{\sigma_{x}}$, where $\sigma_{x}$ is the Pauli $x$ matrix.
	
\begin{figure}[htbp]
		\centering
		\includegraphics[width=\linewidth]{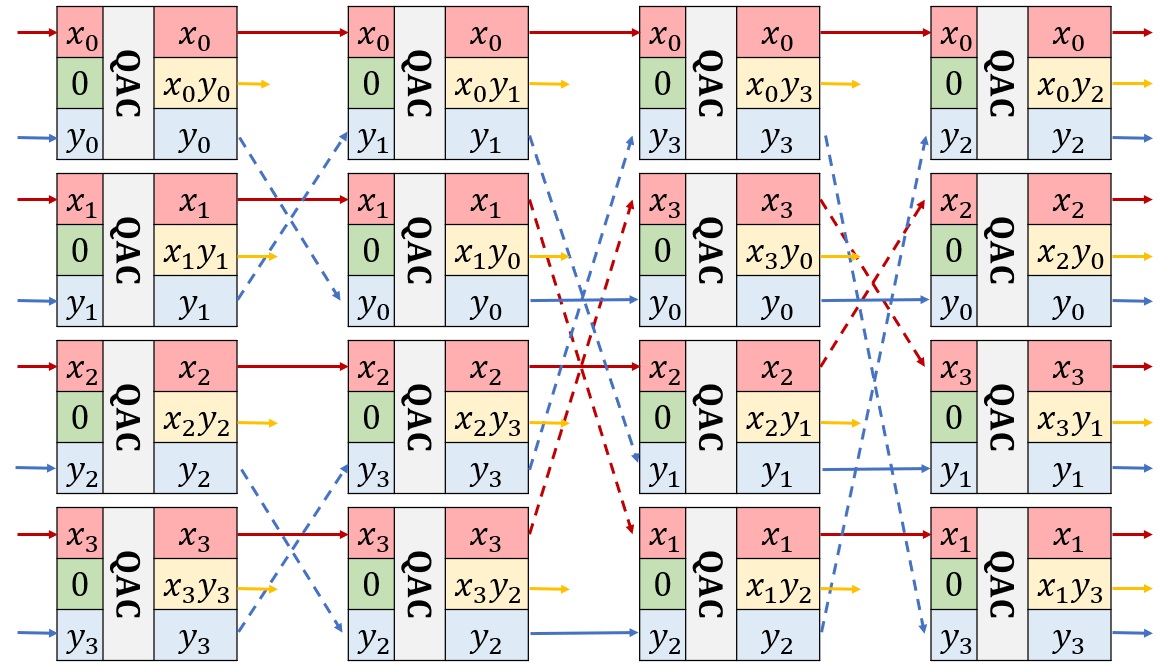}
		\caption{Schematic representation of quantum PPG circuit in multiplication of two $4$-bits numbers}
		\label{PPG}
	\end{figure}

	Using QACs the multiplication of all multiplicand and multiplier qubits are obtained.
	The diagram of the quantum partial product generation circuit of $4 \times 4 $ multiplier using QAC is shown in Fig.\eqref{PPG}.
	
	The outputs of the PPG circuit must be purposefully added together in the PPA circuit to achieve the final result.
	
	\subsubsection{Partial Product Addition circuit (PPA)}
\begin{figure}[htbp]
		\centering
		\includegraphics[width=\linewidth]{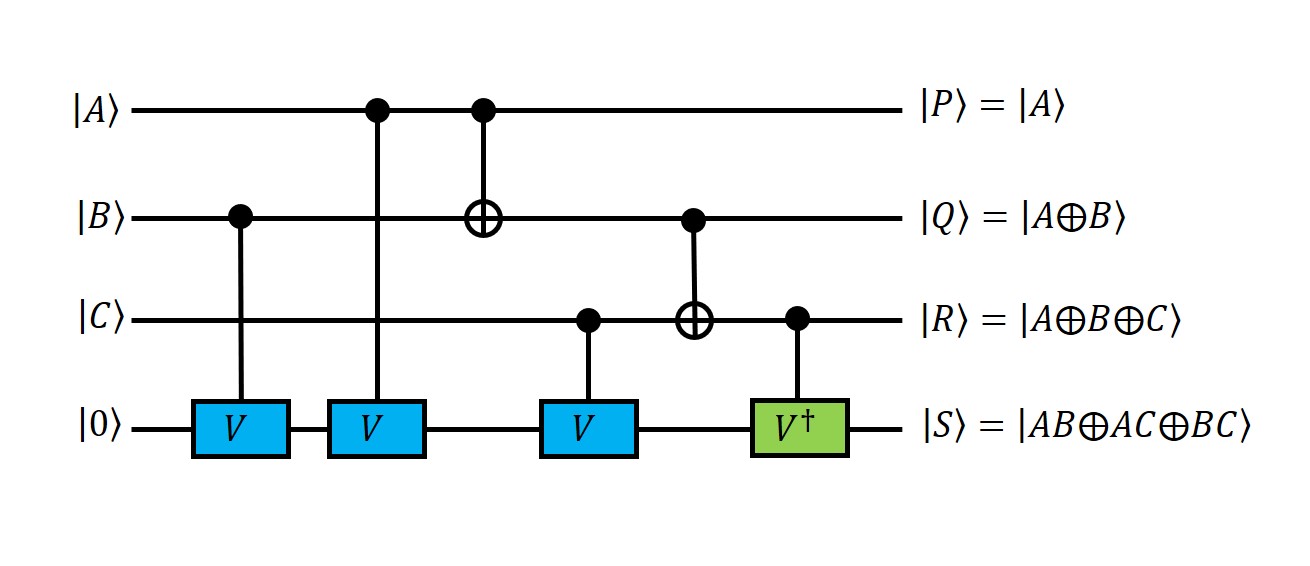}
		\caption{Schematic representation of Quantum Full Adder (QFA) circuit.}
		\label{QFA}
	\end{figure}

	The inputs of the PPA circuit are the outputs of the PPG circuit which must be added purposefully. In order to add two qubits, the quantum full adder circuit (QFA) is used which is shown in Fig.\eqref{QFA}. 
	
\begin{figure}[htbp]
		\centering
		\includegraphics[width=\linewidth]{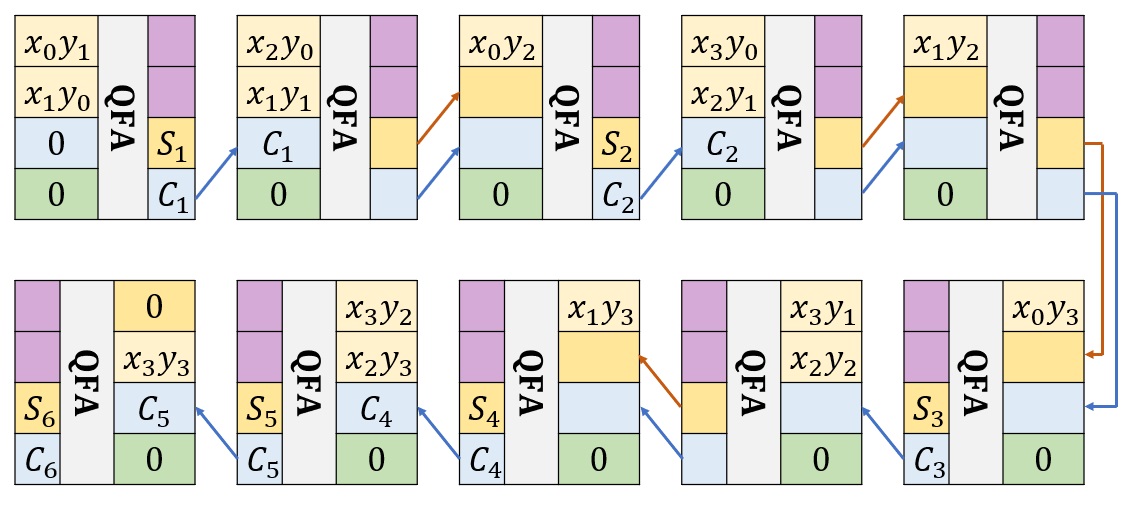}
		\caption{Schematic representation of quantum PPA circuit in multiplication of two $4$-bits numbers}
		\label{PPA}
	\end{figure}

	The QFA circuit adds inputs A and B and C (carry) and shows the sum and carry in output R and S respectively. PPG circut composed of many QFA circuits to get the final result ( $S_0....S_6 $ in Fig.\eqref{Multiply}).
	An example on one arrangement of the QFA circuits to perform the addition operation of $4 \times 4$ multiplier is shown in Fig.\eqref{PPA}.   
	
	After presenting the components of the quantum circuit for the grade-school algorithm, we can now estimate the basic resources required for the circuit, such as: Depth, Cost and Ancillas. As shown in Fig.\eqref{Anding} and Fig.\eqref{QFA}, the depth of each QAC and QFA circuit is $5$ and the cost of QAC and QFA are $5$ and $6$, respectively. Also, each QAC and QFA circuit has one ancilla. By putting these together, we obtain the depth, cost, and ancillas required for multiplying $n$ qubits by $n$ qubits, for different values of $n$, as shown in TABLE \ref{table0}.
	\begin{table}[htbp]
		\begin{tabular*}{\columnwidth}{@{\extracolsep{\fill}}lcccl}
			\hline
			Number Size (in bits)      & $n$ &                      \\ \hline
			Depth                      & $5{n^2} - 5n + 10$      &  \\
			Cost                       & $11{n^2} - 12n + 12$    &  \\
			Ancillas     & $2{n^2} - 2n + 2$          \\ \hline
		\end{tabular*}
		\caption{\label{table0} The estimation of basic resources required for implementing a quantum circuit of the grade-school algorithm.}
	\end{table}
	
	\subsection{quantum circuit of Karatsuba’s multiplication algorithm}
	In this section, we investigate the quantum circuit implementation of Karatsuba’s algorithm \cite{portugal2006reversible}. This algorithm reduces the circuit size by recursively decomposing the multiplication problem of size $n$ into three sub-multiplication problems of size $n/2$ each and achieves an asymptotic speedup over the grade-school algorithm. Actually, the algorithm reduces the number of operations from $T\left(n\right)={n^2}$ in the grade-school method to $T\left(n \right)={n^{\log_2 3}}$, where $T(n)$ is the number of operations for multiplication of $n$-digits numbers. The Karatsuba algorithm consists of the following steps:
	
	\begin{enumerate}
		\item	Input two $n$-digit numbers $X$ and $Y$ (for simplicity, assume that $n$ is a power of $2$).
		\item If $n>1$, then split $X$ and $Y$ into two halves, i.e. $\displaystyle X = 2^{n/2}X_1 + X_2$, $\displaystyle Y = 2^{n/2}Y_1 + Y_2$. Note that $\displaystyle X_1,Y_1,X_2,Y_2$ have $\displaystyle n/2$ digits each.
		\item Compute $\displaystyle U = {Karatsuba}^{[n/2]}(X_1,Y_1)$.
		\item Compute $\displaystyle V = {Karatsuba}^{[n/2]}(X_2,Y_2)$.
		\item Compute $\displaystyle W = {Karatsuba}^{[n/2]}(X_1 + X_2,Y_1 + Y_2)$.
		\item Compute $\displaystyle Z = W - (U + V)$.
		\item Compute $\displaystyle P = 2^nU + 2^{n/2}Z + V$.
		\item Return $P$.
	\end{enumerate}
	
\begin{figure}[htbp]
		\centering
		\includegraphics[width=\linewidth]{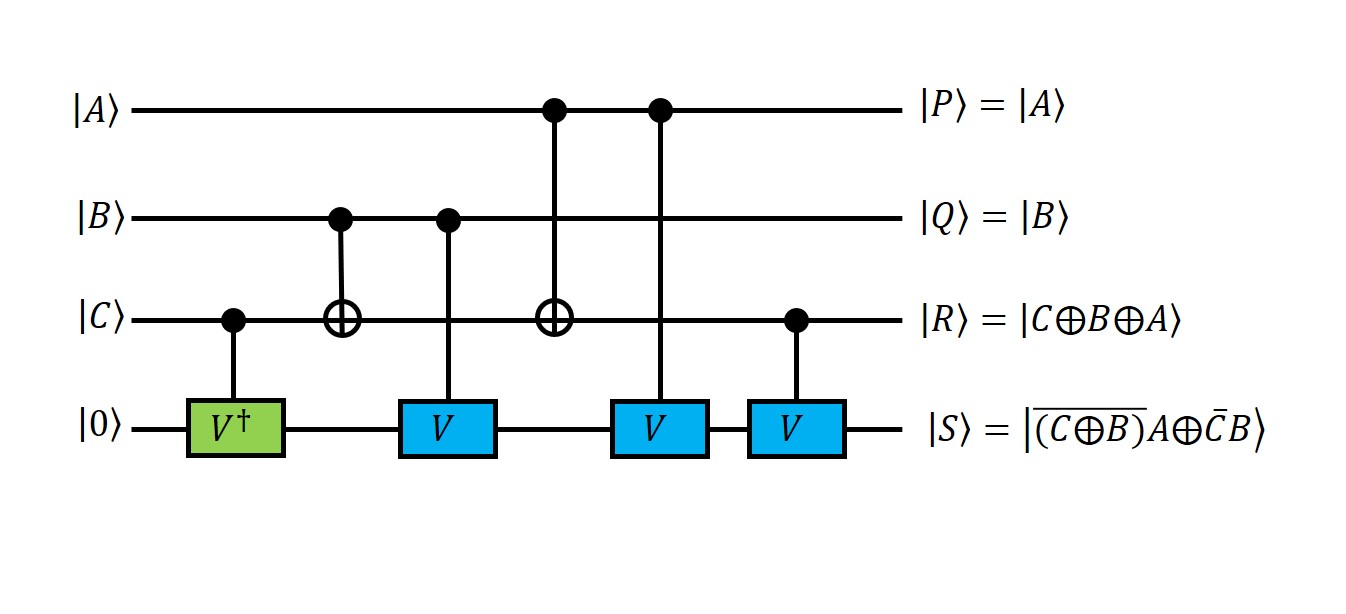}
		\caption{Schematic representation of Quantum Full-Subtractor circuit.}
		\label{M1}
	\end{figure}

	Unlike the grade-school method, Karatsuba’s multiplication algorithm requires not only the Quantum Adding Circuit but also the Quantum Subtraction Circuit as a necessary component. Fig.\eqref{M1} shows a schematic design of an optimized version of the Quantum Full-Subtractor. A Quantum Full-Subtractor circuit has four inputs $(A,B,C,0)$ and four outputs: $(P, Q, R, S)$, in which, the \textit{Difference} and \textit{Borrow} are $R = \left| {C \oplus B \oplus A} \right\rangle$ and $S = \left| {\overline {C \oplus B} A \oplus \overline C B} \right\rangle$, respectively. Note that the cost and the depth of a Quantum Full-Subtractor are 6 and the number of ancillas is 1 \cite{babu}.
	
\begin{figure}[htbp]
		\centering
		\includegraphics[width=\linewidth]{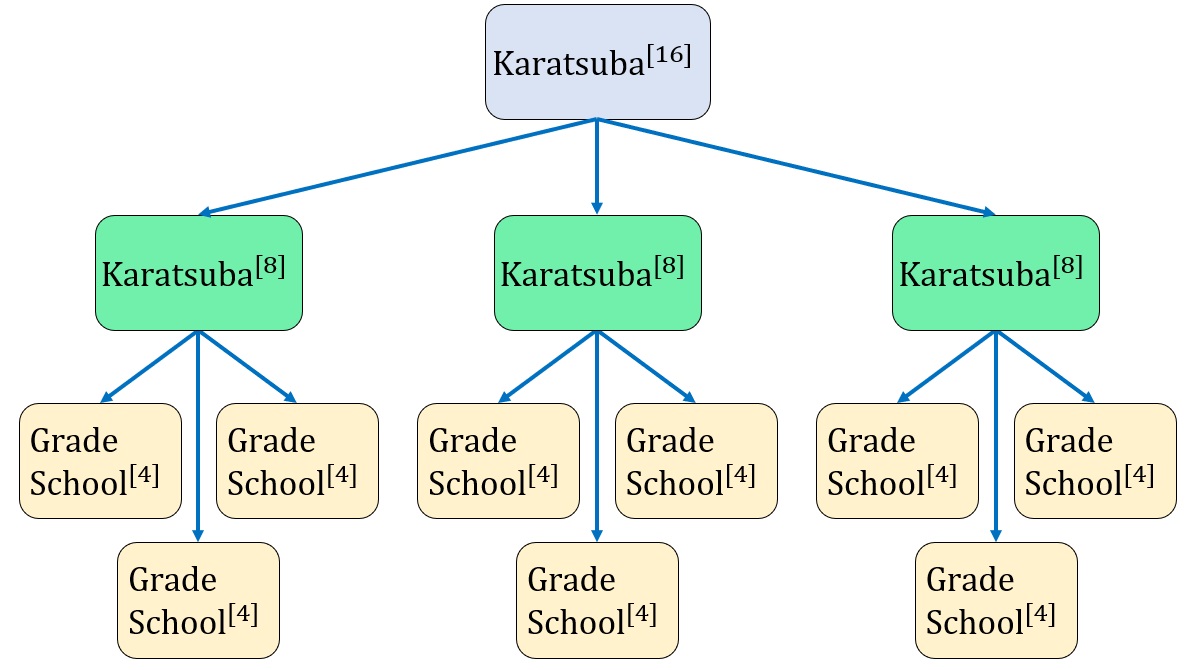}
		\caption{A schematic representation of quantum Karatsuba’s algorithm for 16 qubits.}
		\label{M2}
	\end{figure}

	Fig.\eqref{M2} shows a schematic representation of a simple structure of the Karatsuba multiplication algorithm for 16 qubits. ${Karatsuba}^{[16]}$ is divided into three Karatsuba multiplication algorithms for $8$ qubits each. Each ${Karatsuba}^{[8]}$ is divided into three ${Karatsuba}^{[4]}$ methods, which have almost the same cost as $ {Grade-School}^{[4]}$.
	
\begin{figure}[htbp]
		\centering
		\includegraphics[width=\linewidth]{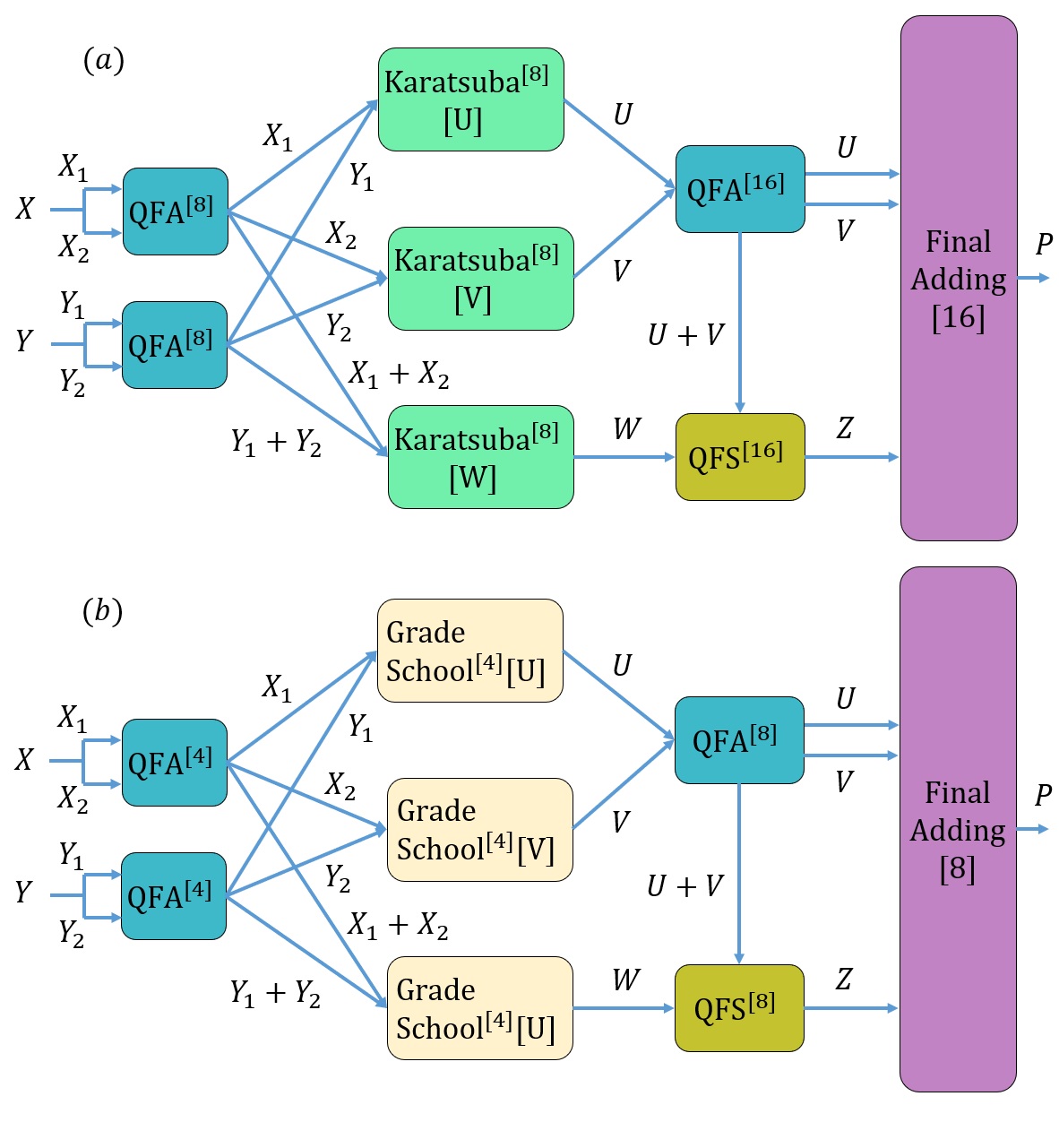}
		\caption{a) Schematic representation of Karatsuba multiplication circuit for $16$ qubits ($Karatsuba{^{\left[ {16} \right]}}$) and b) A detailed view of the $Karatsuba{^{\left[ {8} \right]}}$ circuit, which uses quantum circuits for full adder, full subtractor and Grade-school multiplication of 4 qubits each. The quantum circuits are denoted by $QFA{^{\left[ {4} \right]}}$, $QFA{^{\left[ {4} \right]}}$ and $Grade-School{^{\left[ {4} \right]}}$, respectively.}
		\label{M3}
	\end{figure}

	Fig.\eqref{M3}-a shows how to implement Karatsuba's algorithm for $16$ qubits using quantum circuits. We use the symbols $QFA^{[8]}$, $QFS^{[8]}$ and $Karatsuba^{[8]}$ to represent the quantum circuits for full adder, full subtractor and Karatsuba multiplication of $8$ qubits each. Fig.\eqref{M3}-b zooms in on the circuit of $Karatsuba^{[8]}$, which uses the quantum circuits for full adder, full subtractor and grade-school multiplication of $4$ qubits each, denoted by $QFA^{[4]}$, $QFS^{[4]}$ and ${Grade-School}^{[4]}$.
	
	The purple box in Fig.\eqref{M3} illustrates the quantum circuit of Final Adding which performs the $7$-th step of the algorithm, i.e. $P = {2^n}U + {2^{\frac{n}{2}}}Z + V$. The $Final\,Adding^{[n]}$ module comprises $n$ $QFA^{[1]}$ modules, and therefore requires $n$ ancilla qubits.
	
	After presenting the quantum circuit for the Karatsuba algorithm, we can now estimate the basic resources required for the circuit.
	To do so, we provide the resource estimate for the modules that are used in the algorithm, namely ($QF{A^{\left[ n \right]}},;QF{S^{\left[ n \right]}},Final\,Addin{g^{\left[ n \right]}}$). The TABLE \ref{table1} displays the depth, cost, and ancillas for each of these circuits
	
	\begin{table}[htbp]
		\begin{tabular*}{\columnwidth}{@{\extracolsep{\fill}}lcccl}
			\hline
			Module          & $QF{A^{\left[ n \right]}}$ & $QF{S^{\left[ n \right]}}$ & $Final\,Addin{g^{\left[ n \right]}}$ &  \\ \hline
			Depth                      & 5n                         & 6n                         & 5n                                  &  \\
			Cost  & 6n                         & 6n                         & 6n                                  &  \\
			Ancillas     & n                          & n                          & n                                  &  \\ \hline
		\end{tabular*}
		\caption{\label{table1}The resource estimate of modules used in quantum Karatsuba’s algorithm for $n$ qubits.}
	\end{table}
	
	From Fig.\eqref{M3} one can easily derives that in general, the $Karatsub{a^{\left[ n \right]}}$ has three $Karatsub{l^{\left[ n/2 \right]}}$, two $QF{A^{\left[ n/2 \right]}}$, one $QF{S^{\left[ n \right]}}$, one $QF{A^{\left[ n \right]}}$ and finally one $Final\,Addin{g^{\left[ n \right]}}$. Now, considering the TABLE \ref{table1}, cost, depth, and the number of ancillas of the $Karatsub{a^{\left[ n \right]}}$ algorithm can be derived as follows.
	
	Let $T(n)$ be the resource (depth, cost and ancilla) required by the $Karatsub{a^{\left[ n \right]}}$ algorithm. The following equation gives us a recursive relation for the resource count
	\begin{equation}
		T\left( n \right) = \alpha T\left( {n/2} \right) + O\left( n \right).
	\end{equation}
	In the resource counts for depth, cost, and ancillas, we set $\alpha=3$ for the latter two and $\alpha=1$ for the former. The additional $O(n)$ term accounts for the resources required by other modules, such as $QF{S^{\left[ n \right]}}$, $QF{A^{\left[ n \right]}}$, and so on.
	
	The estimated resources for the Karatsuba algorithm are listed in TABLE \ref{table2}, obtained by evaluating $O(n)$ for different resource values and solving the corresponding recursive equation while taking into account the relevant initial conditions.
	
	\begin{table}[htbp]
		\begin{tabular*}{\columnwidth}{@{\extracolsep{\fill}}lcccl}
			\hline
			Number Size (in bits)       & $n$                                   \\ \hline
			Depth                       & $37n-78$                              \\
			Cost                        & $\frac{332}{9}n^{\log_2{3}}-48n$      \\
			Ancillas                    & $\frac{58}{27}n^{\log_2{3}}-8n$      \\ \hline
		\end{tabular*}
		\caption{\label{table2} The estimation of basic resources required for implementing a quantum circuit of the Karatsuba algorithm.}
	\end{table}
	
	Here, we presented a basic implementation of the Karatsuba algorithm on quantum computers. It's worth noting that several enhancements to this implementation have been explored in the literature. For instance, in reference \cite{parent2017improved}, a significant reduction in space requirements (Ancillas), from $O(n^{1.585})$ to $O(n^{1.427})$, was reported. Additionally, in another reference \cite{gidney2019asymptotically}, the space complexity was further reduced to $O(n)$, while the gate complexity (Cost) remained at $O(n^{1.585})$.
	
	
	\section{Quantum Multiplication Algorithm Based on Convolution Theorem} \label{sec3}
	
	In this section, we present a quantum algorithm for multiplying integers based on the convolution theorem. We first review the classical convolution theorem and its application to integer multiplication. Then, we show how to adapt the convolution theorem to the quantum domain using three key steps: qubit encoding, quantum Fourier transform and element-wise product.We also analyze the advantages of our algorithm over classical methods and provide simulation results using Qiskit.
	
	\subsection{Convolution Theorem} \label{sub3a}
	The Convolution Theorem is a fundamental concept in signal processing and mathematics that relates the convolution operation in the time domain to multiplication in the frequency domain. It provides a powerful tool for analyzing and manipulating discrete vectors in various domains.
	
	In the context of discrete signals represented as vectors, the convolution operation combines two vectors to produce a third vector that represents the interaction between the two. Mathematically, the convolution of two discrete vectors $f[j]$ and $g[j]$ is defined as
	\begin{equation}\label{eq:convolution}
		(f*g)[j] = \sum_{i=1}^{D} f[i]g[j-i],
	\end{equation}
	where $*$ denotes the convolution operator, $D$ is the length of the vector, and when $j-i$ is negative, we wrap around the indices by adding the length of the vector to ensure they fall within the valid range. Mathematically, if $j - i < 0$, then we can express the wrapped index as $(j - i)\,\text{mod}\,D$. This operation computes the sum of the element-wise product of the two vectors, with one vector ($g[j]$) being time-reversed and shifted before multiplication.  
	
	The Convolution Theorem states that the Discrete Fourier Transform (DFT) of the convolution of two vectors in the time domain is equal to the element-wise multiplication of their individual DFTs. In other words, if $F(f)$ and $F(g)$ represent the DFTs of vectors $f[j]$ and $g[j]$, respectively, then the DFT of their convolution $(f * g)[j]$ is given by
	\begin{equation}
		F(f * g) = F(f) \times F(g),
	\end{equation}
	where $\times$ represents the element-wise multiplication of the complex-valued frequency components.
	
	This theorem provides a powerful tool for signal-processing tasks. It allows us to efficiently perform convolutions by simply transforming the vectors to the frequency domain using the Discrete Fourier Transform (DFT), multiplying their spectra, and then transforming the result back to the time domain using the Inverse Discrete Fourier Transform (IDFT). This approach can significantly simplify the computation of convolutions, especially when dealing with large vectors or complex systems.
	
	
	
	\subsection{Integers Multiplication using Convolution Theorem} \label{sub3b}
	The Convolution Theorem is a powerful mathematical tool that can be used to perform integer multiplication efficiently. Instead of using the traditional grade-school multiplication algorithm, which can be slow for large integers, we can convert the integers to vector representations and then perform convolution on the digit sequences of the vectors. By using the Convolution Theorem, we can obtain the multiplication value of the two integers as a vector that can be converted back to an integer representation. This approach offers an efficient and mathematically sound method for performing integer multiplication.
	
	To convert an integer to a vector representation, we can use the binary representation method. This involves dividing the integer by 2 repeatedly and noting down the remainders. The resulting remainders form the binary digits of the integer. For example, the integer $27$ can be converted to the binary representation $11011$. Each digit in the binary representation corresponds to an element in the vector. Therefore, we can represent the integer $27$ as the vector $(1, 1, 0, 1, 1)$.
	
	When performing the multiplication of two $n$-digit integers, the resulting product will have $2n$ digits. To accommodate this, we need to ensure that the vectors representing the integers are zero-padded. This means adding zeros to the most significant side of the vectors, extending their size to $2n$ digits or elements. By zero-padding the vectors, we create enough space for the resulting multiplication to fill in the expanded vector. This step is crucial to ensure the correct representation of the product and to preserve the accuracy of the multiplication operation. By performing zero-padding, one can verify that integer multiplication is equivalent to the convolution of their corresponding vectors.
	
	The time complexity of integer multiplication using the Convolution Theorem is significantly improved compared to traditional methods. The key factor contributing to this improvement is the utilization of fast algorithms for the Discrete Fourier Transform (DFT) and Inverse Discrete Fourier Transform (IDFT). The DFT and IDFT can be computed efficiently using algorithms such as the Fast Fourier Transform (FFT) and its inverse. These algorithms have a time complexity of $O(n \log_2 n)$, where $n$ is the size of the vectors representing the integers. Since the size of the vectors is $2n$ (to accommodate the $2n$-digit product), the time complexity of the overall integer multiplication using the Convolution Theorem is $O(n \log_2 n)$. This is a significant improvement compared to the $O(n^2)$ time complexity of traditional long multiplication algorithms. The Convolution Theorem, combined with fast Fourier transform algorithms, offers a faster computational approach for integer multiplication.
	
	While the Convolution Theorem provides an efficient method for integer multiplication using vector convolution, it is important to note that directly implementing this method on a computer may not be practical. One reason is that the Discrete Fourier Transform (DFT) and its inverse, which are key components of the Convolution Theorem, produce complex numbers as intermediate results. Storing and manipulating complex numbers on a computer typically requires a floating-point architecture. However, working with floating-point numbers can introduce rounding errors and loss of precision, which can impact the accuracy of the multiplication results. As a result, direct application of the Convolution Theorem for integer multiplication on a computer may not be suitable, especially when high precision and accuracy are required.
	
	In response to the limitations of complex numbers and floating-point arithmetic in the Convolution Theorem for integer multiplication, alternative methods have emerged. The use of the Discrete Fourier Transform (DFT) over Finite Fields has shown promise, enabling integer multiplication without rounding errors or precision loss. Notably, Strassen introduced this approach, achieving a time complexity of $O(n \log_2 n \log_2 \log_2 n)$. By combining the Convolution Theorem with the DFT over Finite Fields, accurate and efficient multiplication of integers is made possible.
	
	Quantum mechanics can offer significant improvements to multiplication algorithms that utilize the Convolution Theorem in two key ways. Firstly, by leveraging the principles of quantum computing, the binary vector representation of an integer with $n$ binary digits, which would typically require $n$ classical bits, can be encoded using only $\log_2 n$ qubits. This exponential reduction in qubit usage is due to the fact that the Hilbert vector space of $n$ qubits has a dimension of $2^n$. Secondly, since qubits can represent complex numbers, one can utilize Quantum Fourier Transform (QFT) techniques on the qubits, mitigating the occurrence of rounding errors that are common in classical computations. The inherent properties of quantum systems enable more precise and accurate calculations when employing the Convolution Theorem for integer multiplication.
	
	Furthermore, while quantum mechanics presents opportunities for improving multiplication algorithms based on the aforementioned advantages, there are challenges when it comes to performing element-wise multiplication using unitary operations. It can be shown that exact element-wise multiplication cannot be achieved through unitary operations alone. However, in light of this limitation, we propose a novel approach utilizing a probabilistic circuit for element-wise multiplication. By incorporating probabilistic elements into the quantum circuit, we can overcome the inherent constraints and enable efficient and accurate element-wise multiplication of quantum states. This innovative approach paves the way for leveraging the power of quantum mechanics to enhance multiplication algorithms beyond the limitations of unitary operations.
	
	In what follows, we will explore the methods employed to encode an integer in the Hilbert space of qubits, as well as introduce a novel approach for conducting element-wise multiplication through the utilization of a probabilistic circuit.
	
	
	\subsection{Vector Representation of Integers in Hilbert Space} \label{sub3c}
	
	In this section, we will explore the concept of representing integers as vectors in the Hilbert space. Each integer, denoted as $A$, can be decomposed as $A = \sum_{j=0} A[j] B^j$, where $B$ is referred to as the base. In binary representation, the base is equal to 2. By denoting the powers of 2, such as $2^0$, $2^1$, $2^2$, and so on, with corresponding vectors $(0,...,0,0,1)$, $(0,...,0,1,0)$, $(0,...,1,0,0)$, and so forth, we can represent an integer $A$ as a binary vector $A = (A[n-1],...,A[1],A[0])$, which requires $n$ classical bits for storage.
	
	To efficiently encode the powers of 2 ($2^0$, $2^1$, $2^2$, and so on) in the Hilbert space, we can take advantage of the fact that they share the same base. Consequently, it is possible to encode only the exponents in the Hilbert space. This allows us to represent $2^0$, $2^1$, $2^2$, and subsequent powers of 2 in the Hilbert space using the respective vectors $\ket{0\dots00}$, $\ket{0\dots01}$, $\ket{0\dots10}$, and so on. By employing this encoding scheme, an integer $A$ can be represented in the Hilbert space as the sum $A = \frac{1}{Z}\sum_{j=0} A[j] \ket{j}$, where the index $j$ in $\ket{j}$ represents the binary form of $j$ and, $Z=\sqrt{\sum_{j=0}A[j]^2}$ is the normalization factor. 
	
	Notably, this encoding technique requires only $\log_2 n$ qubits to represent an $n$-digit integer. By utilizing the quantum properties of superposition and entanglement, we can achieve a compact representation of integers in the Hilbert space, leading to potential space savings and computational benefits in certain applications.
	
	With this encoding scheme, it is worth noting that zero-padding in the quantum representation of integers is highly efficient. In the quantum setting, adding an additional qubit doubles the dimension of the Hilbert space. This means that for zero-padding, where additional zeros are appended to the binary representation, only one extra qubit is required. In contrast, in the classical case, zero-padding necessitates doubling the number of classical bits to accommodate the expanded number of digits. This property highlights the advantage of quantum encoding in terms of space efficiency for zero-padding operations.
	
	As indicated in \cite{shukla2023efficient}, the gate complexity needed to encode an $n$-bit number into $\log_{2} n$ qubits, following the described method, is $O(\log_{2} n)$.
	
	
	\subsection{Probabilistic Quantum Circuit for element-wise Multiplication} \label{sub3d}
\begin{figure}[htbp]
		\centering
		\includegraphics[width=\linewidth]{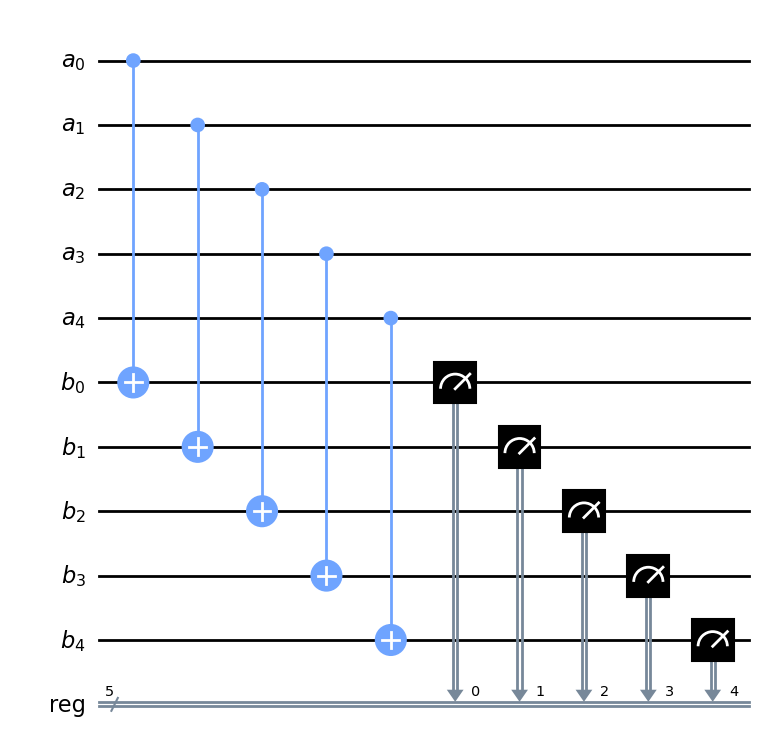}
		\caption{Quantum element-wise multiplication circuit for 5 qubits}
		\label{fig:pairwise_multiplication}
	\end{figure}

	The element-wise multiplication quantum circuit is a circuit that receives two $k$-qubit ($k=\log_2 n$) systems with states ${\ket{\psi_{\text{in}}^{(1)}} = \sum_{i\in\{0,1\}^k}\alpha_{i}\ket{i}}$  and ${\ket{\psi_{\text{in}}^{(2)}} = \sum_{i\in\{0,1\}^k}\beta_{i}\ket{i}}$ as input and puts their element-wise multiplication ${\ket{\psi_{\text{out}}^{(1)}} = \frac{1}{\sqrt{\sum_{j}|\alpha_j\beta_j|^2}} \sum_{i\in\{0,1\}^k}\alpha_{i}\beta_{i}\ket{i}}$ on the first $k$ qubits as output. One can check that the circuit of Fig.\eqref{fig:pairwise_multiplication} satisfies the desired requirement if the result $\ket{00\dots0}$ is obtained after measuring the second $k$ qubits. This figure depicts a quantum element-wise multiplication circuit specifically designed for 5 qubits. However, it is important to note that the generalization of this circuit for $k$ qubits is straightforward. To extend the circuit for $k$ qubits, one can simply include a CNOT gate between any two corresponding qubits of the input registers $a$ and $b$. By applying the CNOT gate between qubits with the same indices, the circuit can be scaled up to accommodate any desired number of qubits, allowing for efficient element-wise multiplication on a quantum system.
	
	The proposed circuit for element-wise multiplication of two $k$-qubits indeed requires $k$ CNOT gates. Importantly, since CNOT gates act on different qubits, they can be applied simultaneously in a single clock cycle. This parallel application of the CNOT gates allows for efficient computation of the element-wise multiplication, reducing the overall computational time required. Therefore, by applying all the CNOT gates concurrently, the circuit can perform the multiplication operation swiftly and effectively on a quantum system.
	
	This aspect of the algorithm is inherently probabilistic, with a success probability of $O(1/n)$. To ensure the multiplication algorithm's correct execution, it must be iterated $O(n)$ times. Employing the quantum amplitude amplification method \cite{brassard2002quantum}, it becomes feasible to enhance the success probability, achieving the $\ket{00\dots0}$ state for the second qubits by introducing a unitary operator and repeating it $O(\sqrt{n})$ times (as elaborated in the forthcoming section), akin to the Grover search algorithm. This adjustment results in an almost deterministic circuit.
	
	
	\subsection{Quantum Amplitude Amplification} \label{sub3}
	Amplitude amplification is a fundamental technique in quantum computing and quantum algorithms that allows us to enhance the probability of measuring a desired state while suppressing the probability of measuring other states. It was first introduced by Gilles Brassard and Peter Høyer in 1997 \cite{brassard1997exact} and independently rediscovered by Lov Grover in 1998 \cite{grover1998quantum}. This algorithm provides a quadratic speedup over classical search algorithms for unstructured databases or searching unsorted lists. In the following, we briefly explain this method.
	
	Let's suppose we're dealing with a quantum system's state space, which is represented by a Hilbert space of $N$ dimensions $\mathcal{H}$. This space is constructed using orthonormal computational basis states. Furthermore, suppose we have a Hermitian projection operator, denoted as $P$ which can be utilized to divide the Hilbert space $\mathcal{H}$ into two distinct subspaces that are mutually orthogonal. These subspaces are referred to as the ``good subspace" $\mathcal{H}_{1}$ and the ``bad subspace" $\mathcal{H}_{0}$. We can decompose the state vector $\ket{\psi}\in \mathcal{H}$ into these two subspaces by projection operator $P$ as: $\ket{\psi}=\cos(\theta)\ket{\psi_{0}}+\sin(\theta)\ket{\psi_{1}}$.
	
	Defining operator $S_{P}=1-2P$ which flips the phase of the state in good subspace and operator $S_{\psi}=1-2\ket{\psi}\bra{\psi}$ which flips the phase of state $\ket{\psi}$, we can construct unitary operator $Q(\psi,P)=-S_{\psi} S_{P}$. The effect of operator $Q$ acting on state $\ket{\psi}$, results in a rotation with an angle of $2\theta$.

	Applying the operator $Q$ repeatedly $n$ times on the state $\ket{\psi}$ results in:
	\begin{equation}
		Q^{n}\ket{\psi}=\cos((2n+1)\theta)\ket{\psi_{0}}+\sin((2n+1)\theta)\ket{\psi_{1}}.
	\end{equation}
	
	The process involves transitioning the state between the good and bad subspaces through rotation. After $n$ iterations, the likelihood of locating the system in a favorable state is determined by the function $\sin^{2}((2n+1)\theta)$. To maximize this probability, we should select: $n=[\frac{\pi}{4\theta}]$.  So far, in each step of the process, the technique has been boosting the strength of the favorable states, which is why it's called Amplitude amplification.

	
	\subsection{Qiskit Implementation of Quantum Multiplier Algorithm based on Convolution Theorem} \label{sub3e}
	
	\begin{figure*}[htbp]
		\centering
		\includegraphics[width=0.9\textwidth]{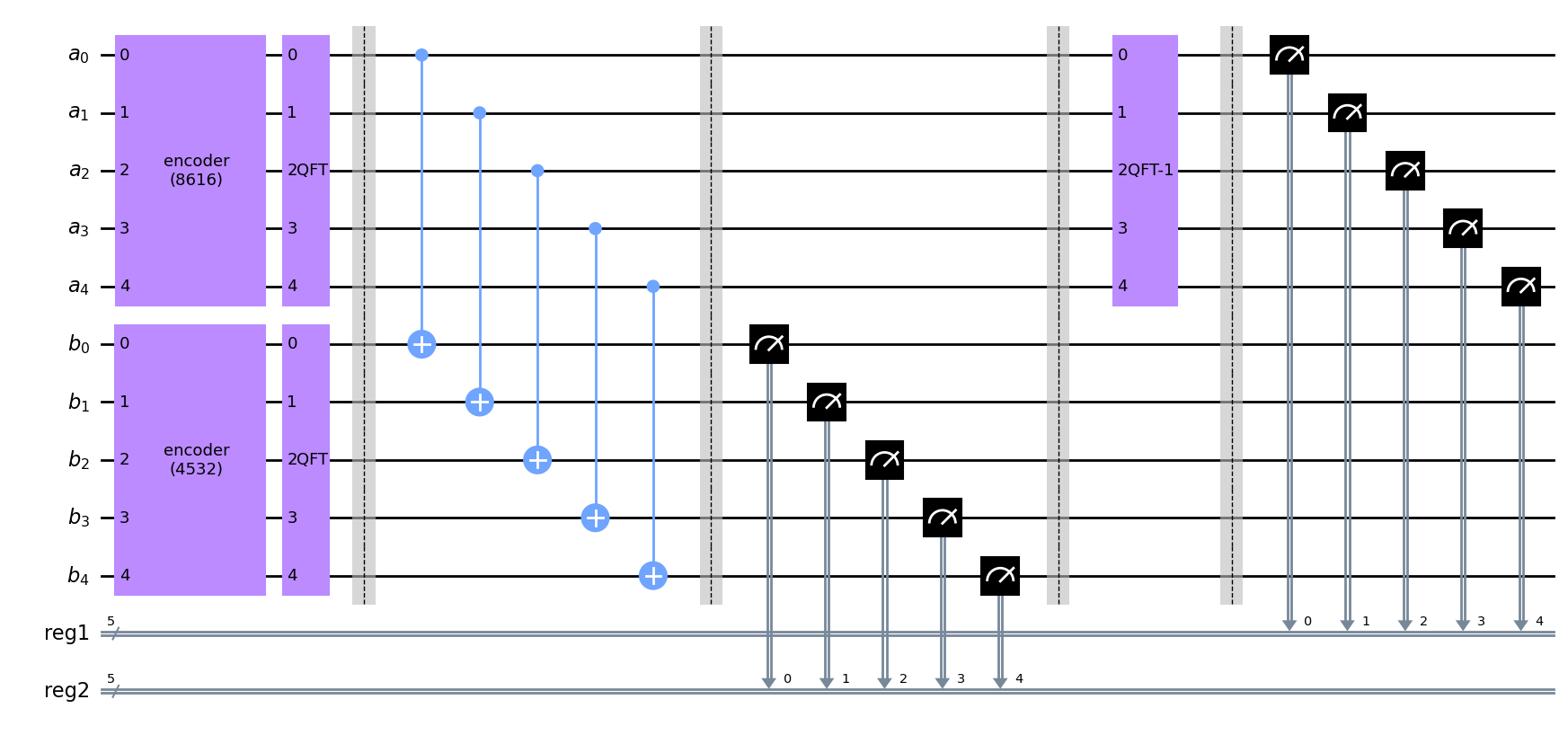}
		\caption{Quantum integer multiplication circuit based on convolution theorem}
		\label{fig:quantum-multiplication-circuit}
	\end{figure*}
	
	In this section, we present a Qiskit implementation of our multiplication algorithm, showcasing its practical application. We aim to multiply two random integers, namely 8616 and 4532, using our proposed encoding method and element-wise multiplication circuit. The Qiskit implementation provides a hands-on demonstration of how quantum computation can be utilized for efficient integer multiplication. By executing the code, the quantum multiplication algorithm will be applied to these specific integers, showcasing the power and potential of quantum computing in tackling real-world computational tasks.
	
	Based on our encoding method, the numbers 8616 and 4532 can be encoded in a 5-qubit system as follows
	\begin{equation}
		\begin{aligned}
			&8616 = 2^3 + 2^5 + 2^7 + 2^8 + 2^{13}\\ &\rightarrow\frac{1}{\sqrt{5}}(\ket{00011}+\ket{00101}+\ket{00111}+\ket{01000}+\ket{01101}),\\
			&4532 = 2^2 + 2^4 + 2^5 + 2^7 + 2^8 + 2^{12}\\ &\rightarrow\frac{1}{\sqrt{6}}(\ket{00010}+\ket{00100}+\ket{00101}+\ket{00111}+\ket{01000}\\
			&+\ket{01100}).
		\end{aligned}
	\end{equation}
	
	\begin{figure}[htbp]
		\centering
		\includegraphics[width=\linewidth]{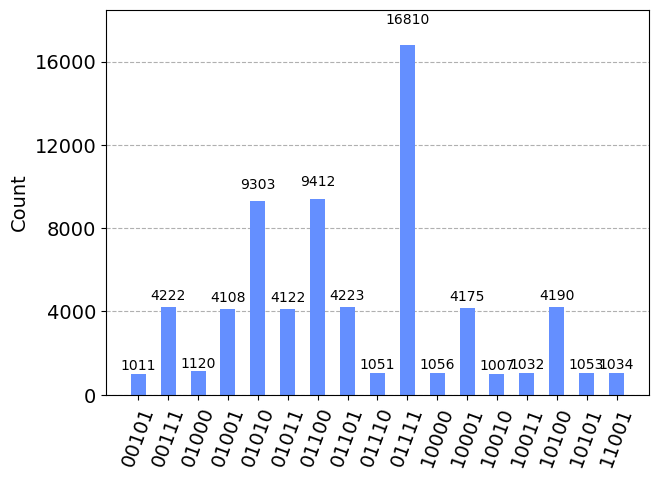}
		\caption{Result of measurement on quantum register $a$}
		\label{fig:quantum-multiplication-result}
	\end{figure}
	
	Fig.\eqref{fig:quantum-multiplication-circuit} displays our proposed quantum multiplication circuit, which is based on the convolution theorem. To assess the circuit's performance, we conducted 1,000,000 runs, and in 68,929 instances, the measurement on quantum register $b$ resulted in the state $\ket{00000}$. Fig.\eqref{fig:quantum-multiplication-result} provides an overview of the measurement results obtained from quantum register $a$. Analyzing these results allows us to infer the state of register $a$ prior to measurement, which is
	\begin{equation}
		\begin{aligned}
			\ket{\psi_{\text{out}}^{(a)}} =& \frac{1}{\sqrt{66}}(\ket{00101}+2\ket{00111}+\ket{01000}+2\ket{01001}\\
			& +3\ket{01010}+2\ket{01011}+3\ket{01100}+2\ket{01101}\\
			& +\ket{01110} + 4\ket{01111} + \ket{10000} + 2\ket{10001}\\
			& +\ket{10010} + \ket{10011} + 2\ket{10100} + \ket{10101}\\
			& +\ket{11001}),
		\end{aligned}
	\end{equation}
	decoding the state obtained from the measurement results of register $a$ yields the product number as
	\begin{equation}
		\begin{aligned}
			&(1\times2^{05})+(2\times2^{07})+(1\times2^{08})+(2\times2^{09})\\
			+&(3\times2^{10})+(2\times2^{11})+(3\times2^{12})+(2\times2^{13})\\
			+&(1\times2^{14})+(4\times2^{15})+(1\times2^{16})+(2\times2^{17})\\
			+&(1\times2^{18})+(1\times2^{19})+(2\times2^{20})+(1\times2^{21})\\
			+&(1\times2^{25}) = 39,047,712.
		\end{aligned}
	\end{equation}
	
\begin{figure*}[htbp]
		\centering
		\includegraphics[width=0.9\textwidth]{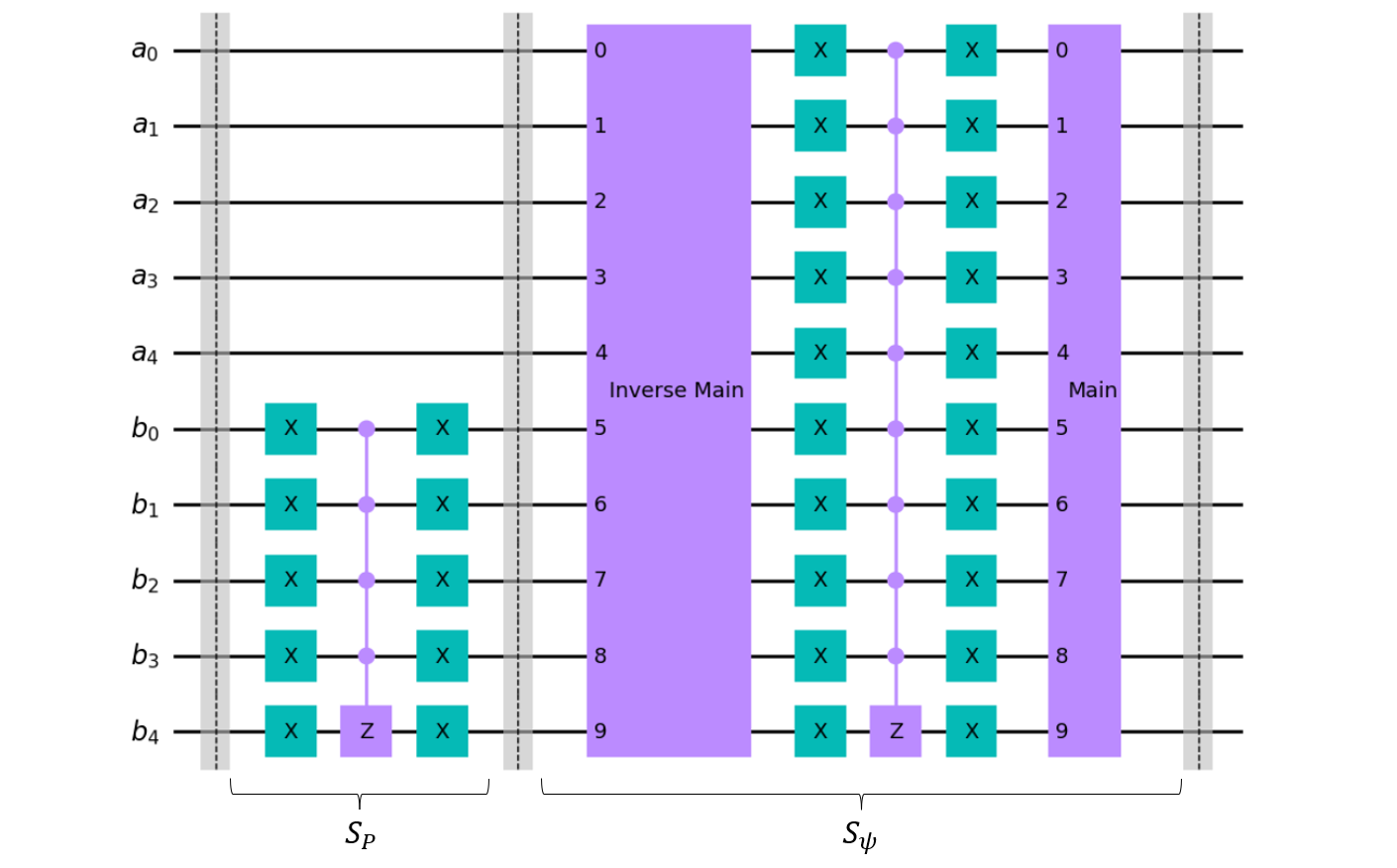}
		\caption{Quantum amplitude amplification gate}
		\label{fig:quantum-amplitude-amplification}
	\end{figure*}

	In the previous example, we didn't utilize the amplitude amplification method. The circuit described there is particularly beneficial for quantum computers with shallow circuit depths. However, for an ideal quantum computer, one can employ the amplitude amplification process to harness computational advantages. We define the \textit{Main} circuit as the concatenation of the following components: Encoder(a), Encoder(b), QFT(a), QFT(b), and CNOT gates. To implement amplitude amplification, a specific quantum gate should be inserted before measuring the quantum register b. This gate needs to be repeated approximately $O(\sqrt{n})$ times for optimal results, as illustrated in Fig.\eqref{fig:quantum-amplitude-amplification}.

	\section{Discussion and Conclusion} \label{sec4}
	Efficient multiplication of large integer numbers is the key component in many fields including cryptography, computer science, and engineering. The optimal design of the algorithm for this task depends on the size and type of the numbers involved, the available resources,  the desired level of accuracy, and hardware constraints.
	
	In the context of quantum computation, multiplication retains its significance as one of the fundamental mathematical operations involved in various quantum algorithms. However, in this context, it can also exploit quantum resources to achieve more efficient performance of the multiplication task. Hence, it is important to propose efficient quantum multiplication algorithms that are suitable for quantum hardware implementation. 
	
	In this paper, inspired by the convolution theorem and motivated by the advantage of QFT over FFT we propose a quantum algorithm for integer multiplication with some advantages using quantum resources. The core of our work was the construction of a quantum version of the convolution theorem that can be implemented with quantum circuits. It consists of three main steps. The first step is to encode the binary vectors of polynomial coefficients into qubits. We find that we only need $\log_2{n}$ qubits to encode a vector for an $n$-bit number. This reduces the space complexity of the algorithm, which is the first benefit we obtain from quantum resources. The second step is to apply the quantum Fourier transform to these vectors, which reduces the time complexity of the algorithm. This is the second benefit we gain from quantum resources. The third step is to construct a quantum circuit for the element-wise product of two vectors in Hilbert space. However, there is no deterministic quantum circuit for this task \cite{lomont2003quantum}. We have to use a probabilistic quantum circuit instead, which increases the time (or space) complexity of the algorithm. The probabilistic quantum circuit exhibits a success probability of $O(1/n)$, while the time complexity of the remaining steps in the algorithm amounts to $O(\log^{2}_{2}{n})$ for each successful execution. Consequently, the overall time complexity of the algorithm is $O(n\log^{2}_{2}{n})$. However, by employing the quantum amplitude amplification method, we can significantly enhance the algorithm's efficiency, reducing its complexity to $O(\sqrt{n}\log^{2}_{2}{n})$. This remarkable improvement surpasses the performance of the most advanced classical algorithm known to date, namely the Harvey algorithm, which operates with a time complexity of $O(n \log_{2} n)$.
	
	At the end, to evaluate the performance of our algorithm, we compare it with other quantum multiplication circuits presented in the paper. We use the basic resources required for each circuit as the comparison metric. TABLE \ref{table4} summarizes the results.
	
	\begin{table}[htbp]
		\begin{tabular*}{\columnwidth}{@{\extracolsep{\fill}}lcccl}
			\hline
			Algorithm & Grade-school                        & Karatsuba                                             & Our Algorithm   \\ \hline
			Depth     & $ \scriptstyle 5{n^2} - 5n + 10$    & $ \scriptstyle 37n-78 $                               & $ O(\sqrt{n}\log^{2}_{2}{n}) $   \\
			Cost      & $ \scriptstyle 11{n^2} - 12n + 12$  & $ \scriptstyle \frac{332}{9} n^{\log_2{3}}-48n $      & $ O(\sqrt{n}\log^{2}_{2}{n}) $   \\
			Ancillas  & $ \scriptstyle 2{n^2} - 2n + 2$     & $ \scriptstyle \frac{58}{27} n^{\log_2{3}}-8n $       & $ \scriptstyle 0 $  \\ \hline
		\end{tabular*}
		\caption{\label{table4} Basic resources of quantum circuits for multiplication of two $n$-bits integers. We compare the depth, cost, ancillas of three algorithms: Grade-School, Karatsuba, and our algorithm.}
	\end{table}
	
	The efficient utilization of quantum resources is essential for the development of quantum algorithms that can outperform classical algorithms. In the context of quantum multiplication algorithms, the efficient implementation of the Fourier transform is a critical component that can significantly impact the performance of the algorithm. In our proposed algorithm, we used the Quantum Fourier Transform (QFT) to perform the Fourier transform operation. While the QFT algorithm is widely used in quantum computing, there are other fast quantum Fourier transform algorithms that can be used to perform the Fourier transform operation \cite{asaka2020quantum,nam2020approximate}. These algorithms have different resource requirements and can potentially improve the performance of the algorithm. Therefore, in future works, it is important to explore and compare the resource requirements of different fast quantum Fourier transform algorithms to optimize the performance of quantum multiplication algorithms.
	
	
	
	\begin{acknowledgments}
		This work was supported by the Research Centre for Quantum Engineering and Photonics Technology, Sharif University of Technology, through the Quantum Algorithm Project under Grant No. 140200401. We would also like to give special thanks to Mahdi Shokhmkar and Diba Masihi for their useful discussions.
	\end{acknowledgments}

	\nocite{*}
	
	\bibliography{references}
	
\end{document}